\nonstopmode
\documentclass[a4paper,12pt]{article}
\usepackage[T1]{fontenc}
\usepackage{amsfonts}
\usepackage{geometry}
\usepackage[utf8]{inputenc}
\usepackage[pdftex]{graphicx}
\usepackage{graphicx}
\usepackage{amsmath,amsthm}
\usepackage[colorlinks=true]{hyperref}
\usepackage{amssymb}
\usepackage{color}
\usepackage{cellspace}
\usepackage{slashbox}
\usepackage{textcomp}
\usepackage{algorithm}
\usepackage{algorithmic}
\usepackage[toc,page]{appendix}
\newtheorem{rque}{Remark}
\newtheorem{de}{Definition}

\newtheorem{hyp}{Assumption}
\newtheorem{lemma}{Lemma}

\newcommand{\mybf}[1]{\mbox{\boldmath{$#1$}}}

\newcommand{\N}{\mathbb N}
\newcommand{\R}{\mathbb R}
\newcommand{\Z}{\mathbb Z}

\newcommand{\Cov}{\mathbf{Cov}}
\newcommand{\Var}{\mathbf{Var}}
\newcommand{\E}{\mathbf{E}}

\everymath{\displaystyle\everymath{}}
\newcommand{\Tint}{T^\mathrm{int}}
\newcommand{\Text}{T^\mathrm{ext}}
\newcommand{\Tabove}{T^\mathrm{above}}
\newcommand{\Tbelow}{T^\mathrm{below}}
\newcommand{\Tcor}{T^\mathrm{cor}}
\newcommand{\Toff}{T^\mathrm{off}}

\makeatletter
\def\@captype{table}
\makeatother

\begin{document}
\date{}
%\maketitle
\title{\textbf{Pick and freeze estimation of sensitivity indices for models with dependent and dynamic input processes}}
\footnotetext[1]{G2ELab, Universit\'e de Grenoble, 38402 St-Martin d'H\`eres, France}
\footnotetext[2]{Laboratoire de Math\'ematiques d'Orsay, B\^atiment 425, Universit\'e Paris-Sud, 91405 Orsay, France}
\footnotetext[3]{GSCOP,46 avenue Félix Viallet, 38031 Grenoble Cedex 1, France}
\author{\renewcommand{\thefootnote}{\arabic{footnote}}
%Didier Dacunha-Castelle\footnotemark[2],
Mathilde Grandjacques\footnotemark[1], 
Alexandre Janon\footnotemark[2],
Beno\^it Delinchant\footnotemark[1],
Olivier Adrot\footnotemark[3]}
\maketitle
%\tableofcontents

\section*{Abstract}

This paper addresses sensitivity analysis for dynamic models, linking dependent inputs to observed outputs. The usual method to estimate Sobol indices are based on the independence of input variables. We present a method to overpass this constraint when inputs are Gaussian processes of high dimension in a time related framework. Our proposition leads to a generalization of Sobol indices when inputs are both dependant and dynamic. The method of estimation is a modification of the Pick and Freeze simulation scheme. First we study the general Gaussian cases and secondly we detail the case of stationary models. We then apply the results to an example of heat exchanges inside a building.

\section{Introduction}
To study physical phenomena, it is useful to build mathematic models which translate them. These models can be used for purposes such as managements or forecasts for example. So it is important for the practitioner to assess the fiability of the models used. The sources of uncertainties in a model may be located at two levels :
\begin{itemize}
\item on the parameters when they are estimated them for example 
\item on the inputs of the model (error of measures, variability of the inputs,$\dots$)
\end{itemize} 
Sensitivity analysis can help to do this work. It aims to quantify uncertainties of each factor on the output of the model. The interest can notably be to :
\begin{itemize}
\item reduce variability of the output
\item prioritize factors : see which factor is the most influent on the output and need more precision on its estimation or its measure
\item calibrate the least influent factors.
\end{itemize}
Among the tools available in global stochastic sensitivity analysis (see for example \cite{saltelli2000sensitivity} and references therein), the most used one is Sobol index defined if the variables are assumed to be independent random variables. Their probability distributions account for the practitioner's belief in the input uncertainty. This turns the model output into a random variable, whose total variance can be split down into different partial variances (this is the so-called Hoeffding decomposition, also known as functional ANOVA, see \cite{liu2006estimating}). Each partial variance is defined as the variance of the conditional expectation of the output with respect to each input variable.  By considering the ratio of each partial variance to the total variance, we obtain the Sobol sensitivity index of the variable \cite{sobol1993,sobol2001global}.
This index quantifies the impact of the variability of the factor on the output.  Its value is between 0 and 1 allowing to prioritize the variables according to their influence. 

Even when the inputs are not independent, it seems reasonable to consider the same Sobol index but with a quite different interpretation. Several approaches have been proposed in the literature about dependent inputs. In their introduction, Mara et al. \cite{mara2011variance} cite some of them, which are claimed to be relevant only in the case of a linear model. In that paper, the authors introduce an estimation method for the Sobol index but this method seems computationally intricate. On the other hand, Kucherenko et al.  \cite{kucherenko2012estimation} rewrite, as we will do, the Sobol index as a covariance between the output and a copy of the output. Another method (\cite{chastaing2012generalized}) modifies the Sobol index definition, which leads to indices that are hard to estimate, as well as results that may seem counter-intuitive (for instance, the indices may not be between 0 and 1). 

Few works propose to study the sensitivity to dynamic inputs. The sensitivity is calculated at each time step $t$ without taking into account the dynamic behaviour of the input. Indeed, the impact of the variability is not always instantaneous. It seems necessary to develop a new method to dynamic dependent inputs. In this way, the Sobol index definition is modified.

We set ourselves in a time related framework and we study the following scalar output $Y$ :
\begin{equation}
Y_t = f_t \left((\mybf{U}_s)_{0\leq s\leq t} \right) , \;\;\; t\in \N
\label{e:model:general}
\end{equation}
The input is a vectorial Gaussian process $(\mybf{U}_s)_{s\in\N} \subset \R^p $. In this context, the sensitivity is defined for $Y_t$ with respect to the input process $(U^1_s)_{0\leq s\leq t}$ (for example). Thus the sensitivity changes with time $t$. The dynamic framework is the most useful in stationary or almost stationary cases. In non stationary cases the problem is no more that a sequence of finite dimensional situations. We focus on two cases with $\mybf{U}_t$ a stationary process : \begin{itemize}
\item $ Y_t= f(\mybf{U}_{t},\mybf{U}_{t-1},\dots,\mybf{U}_{t-M})$ for some $M$
\item $Y_t= f(\mybf{U}_{t},\mybf{U}_{t-1},\dots,\mybf{U}_{0},0,\dots,0)$ deduced from a stationary process given by\\ $Y^{\star}_t= f(\mybf{U}_{t},\mybf{U}_{t-1},\dots,\mybf{U}_{0},\mybf{U}_{-1},\dots)$. \\
This case includes models associated to recurrence equations as Euler schemes of stochastic differential equations.
\end{itemize} 

The method of estimation that seems best suited for functional multidimensional models is the Pick and Freeze scheme  (see \cite{sobol2001global,gamboa2013statistical}). It allows flexibility in the form of the inputs and doesn't care of the number of variables by which it is desired to condition the variance, the only constraint being the assumption of independent inputs. In SPF (Scheme Pick and Freeze), a Sobol index is viewed as the correlation coefficient between the output of the model and its pick-freezed replication. This replication is obtained by holding the value of the variable of interest (frozen variable) and by sampling the other variables (picked variables). The sampled replications are then combined to produce an estimator of the Sobol index.

In a first part, after reminding the definition of Sobol index and the Pick and Freeze scheme, we introduce the definition of the index in a dynamic case. We show that under the hypothesis of Gaussian inputs, it is possible to reduce this problem to the case of independent variables and apply the method Pick and Freeze.
In the second part we present the properties of our index  when inputs are stationary. Finally an application to a physical problem is presented in the last section.

\section*{Notations}
Let us give some notations : \begin{itemize}
\item $X,Z$ random variables
\item $\mybf{U},\mybf{Z}$ random vectors 
\item $\left(\mybf{U}_t\right)_{t\in \N}$ a vectorial process, $dim(\mybf{U}_t)=p$
\item $\mybf{U}_{\lfloor a,b \rfloor}= \left\lbrace \mybf{U}_s, a\leq s \leq b \right\rbrace$,  $p\times(b-a+1)$ matrix, with $-\infty \leq a \leq b \leq +\infty$  
\item $\mybf{U}^*$ or $\mybf{U}_t^*$ is the transposed vector of $\mybf{U}$ or the vectorial process $\mybf{U}_t$
\end{itemize}
If $X_t$ and $\mybf{Z}_t$ are two stochastic vectorial centered processes with $dim(X_t)=1, \; dim(\mybf{Z}_t)=p-1$, we define different covariance matrices as following :
\begin{de}$\gamma_{s,v}^{X Z^j}= \E( X_s Z^j_v)$ the covariance between $X_s$ and $Z^j_v$ where $j$ denotes the jth component of the vector $\mybf{Z}_v$ \end{de}
\begin{de} $\mybf{\gamma}_{\lfloor 0,t \rfloor,v}^{X Z^j}= \E( X_{\lfloor 0,t \rfloor} Z^j_v)$ a $(t+1)$ vector process of generic term $\gamma_{s,v}^{X Z^j}, \;\; 0\leq s \leq t$ \end{de}
\begin{de}\label{def:matrice:Gamma} $\Gamma_{\lfloor 0,t \rfloor,v}^{X Z}$ the $(t+1)\times (p-1)$ covariance matrix of generic term $\mybf{\gamma}_{\lfloor 0,t \rfloor,v}^{X Z^j}$ for $1\leq j \leq p-1, \; 0\leq s \leq t$\end{de}
\begin{de} $\Gamma_{\lfloor 0,t \rfloor,\lfloor 0,u \rfloor}^{X Z}$ the $(t+1)\times (p-1)(u+1)$ matrix using matrix blocks $\Gamma^{X Z}_{\lfloor 0,t \rfloor,v}$ with $0\leq v \leq u $\end{de}

To simplify the exposition we consider an input vector $\mybf{U}=(U^1, \dots, U^2)$. We denote by $X=U^1$ and $\mybf{Z}= \left( U^2,\dots,U^p \right)$ when we are in static context and $X_t=U_t^1$ and $\mybf{Z}_t= \left( U_t^2,\dots,U_t^p \right)$ in the dynamic case.

\section{Sobol indices : extended definition and estimation}
\subsection{Definition in a dynamic context}

We consider the model given by : $Y=f(\mybf{U})$, $\mybf{U}\in\R^p$ is a random vector with known distributions.
We assume that all coordinates of $\mybf{U}$ and $Y$ have a finite non zero variance.

The Sobol index with respect to $X$ is defined by \cite{sobol1993} :
\begin{equation}\label{e:defsob} 
S^{X} = \frac{\Var\left( \E(Y|X)\right) }{\Var (Y)}.
\end{equation}

$S^{X}$ is the Sobol index with respect to $X$. More generally, $S^J$ is the closed Sobol index with respect to the group $J$ of variables $\mybf{U}^J=\left( U^j, j \in J \right)$ :
\begin{equation}
S^{U^J} = \frac{\Var\left( \E(Y|\mybf{U}^J)\right) }{\Var (Y)}.
\end{equation}

 Total indices and higher-order Sobol indices can also be written by taking the sum or the difference of closed indices. Hence, we can restrict ourselves to the case of two (possibly vector) inputs in the model. 

We now introduce the time dimension and we define the input-output relation :
\begin{equation}
Y_t=f_t((\mybf{U}_s)_{0\leq s\leq t})
\end{equation}
 $ (\mybf{U}_s)_{s\in\N} \subset \R^p$ is a vector-valued stochastic process, $f_t$ being a sequence of functions which will be detailed later. Let $\mybf{U}_t = (X_t, \mybf{Z}_t)$ with $X_t=U^1_t$ and $\mybf{Z}_t=(U^2_t,\dots,U^p_t)$.
%(i.e., for each $t\in\N$, $X_t$ is independent from $X_s$ for $s>t$).

For each $t \in \N$, we define a measure of the sensitivity of $Y_t$ with respect to $X_{\lfloor 0,t \rfloor}=(X_0,\dots,X_{t-1},X_t)$ by :

\begin{equation}
\label{Popsi_def}
S^{X}_t = \frac{\Var \left(\E\left(Y_t|X_{\lfloor 0,t \rfloor}\right)\right)}{\Var (Y_t)}
\end{equation}

The index $t \mapsto S^{X}_t$ is called the \emph{projection on the past sensitivity index with respect to $X$}. We notice that, at any given time $t$, we consider the sensitivity of $Y_t$ with respect to all the past $X_{\lfloor 0,t \rfloor}$  values of the $X$ process, not just its value $X_t$.

Of course, the conditional expectation with respect to $X_{\lfloor 0,t \rfloor}$ takes into account the dependence of $\mybf{Z}_{\lfloor 0,t \rfloor}$ with respect to $X_{\lfloor 0,t \rfloor}$.

\begin{rque}
When the inputs are dependent we keep the property that $S^{X}_t \leq 1 $ for any $t$. The classical term of interaction $S^{XZ}$ is not defined \cite{chastaing2011generalized}.
\end{rque}
\subsection{Estimation of $S^X$, the Pick and Freeze method in the independent case}

There exists many methods for estimating $S^X$. One of them is the so-called Pick and Freeze scheme \cite{sobol2001global,sobol1993}. In this case $f$ plays the role of a black box allowing to simulate the input-output relationship without any mathematical description. This method is based on the following lemma \cite{sobol2001global} :

\begin{lemma}{Sobol : }
Let $\mybf{U}=(X,\mybf{Z})$. If $X$ and $\mybf{Z}$ are independent : \[ \Var(\E(Y|X))= \Cov(Y, Y^X) \]
with  $Y^X = f(X,\mybf{Z}'), Y= f(X,\mybf{Z})$ where $\mybf{Z}'$ is an independent copy of $\mybf{Z}$.
\end{lemma}
We can deduce the expression of $S^X$ when $X$ and $\mybf{Z}$ are independent :
\begin{equation}\label{e:pickfreeze} S^X = \frac{\Cov(Y,Y^X)}{\Var (Y)}, \end{equation}

A natural estimator consists in taking the empirical estimators of the covariance and of the variance. Let a $N-$sample $\{ (Y^{(1)},Y^{X,(1)}),\dots,(Y^{(N)},Y^{X,(N)}) \} $ a natural estimator of $S^X$ is :

\begin{equation}
\label{e:esti2}
\widehat S^X=\frac{\frac 1 N \sum_{i=1}^N Y^{(i)} Y^{X,(i)} - (\frac 1 N \sum_{i=1}^N  Y^{(i)} )(\frac 1 N \sum_{i=1}^N  Y^{X,(i)} )}{ \frac{1}{N} \sum_{i=1}^N (Y^{(i)})^2 -  (\frac 1 N \sum_{i=1}^N  Y^{(i)})^2 }
\end{equation}

If $X$ and $\mybf{Z}$ are finite dimensional random vectors this formula can be justified by asymptotic properties when $N \rightarrow \infty$.
The speed of convergence of this estimator is in $O(1/\sqrt{N})$, see Janon et al. \cite{janon2012asymptotic}. In practice it can be approximated by $\frac{C}{\sqrt{N}}$ where  $C$ can be large as we will see later.

In the dependent case the estimation of $S^X_t$ by a Monte-Carlo method is a challenging task, as one cannot be chosen $X=(X_{\lfloor 0,t \rfloor})$ and $\mybf{Z}=(\mybf{Z}_{\lfloor 0,t \rfloor})$ in \eqref{e:pickfreeze} since $X$ and $\mybf{Z}$ are not independent. However, we will see, in the following Section, that, in a particular Gaussian case, whatever the covariance structure is, an efficient Pick and Freeze scheme may be built.

\subsection{Reduction to independent inputs for Gaussian processes}
Suppose that we are able to get another expression of the $Y_t$ output of the type :
\begin{equation}
\label{e:function:g}
Y_t=g_t(X_{\lfloor 0,t \rfloor}, \mybf{W}_{\lfloor 0,t \rfloor})
\end{equation}  
%where $(\bar{X}_s)_{s\leq t}$ is $(X_s)_{s\leq t}$ measurable, 
where $\mybf{W}_{\lfloor 0,t \rfloor}$ is a stochastic process independent of $X_{\lfloor 0,t \rfloor}$, $\mybf{W}_{\lfloor 0,t \rfloor}$ being $ (X_s,\mybf{Z}_s)_{s\leq t}$ measurable.

Then :
\begin{equation}
S^X_t = \frac{\Var\left( \E\left(Y_t=g_t(X_{\lfloor 0,t \rfloor}, \mybf{W}_{\lfloor 0,t \rfloor})|X_{\lfloor 0,t \rfloor}\right)\right)}{\Var (Y_t)}
\end{equation}
is defined as in the classical case of independence.

We now prove that if $Y_t=f_t(\mybf{U}_t,\mybf{U}_{t-1},\dots, \mybf{U}_0)$ there exists $g_t$ satisfying \eqref{e:function:g}.

Let : 
\begin{align*}
&\tilde{\mybf{X}}_t = \E\left(\mybf{Z}_t | X_{\lfloor 0,t \rfloor}\right) = \Lambda X_{\lfloor 0,t \rfloor} \mbox{ for a matrix } \Lambda \mbox{ with } dim(\Lambda)= (p-1) \times (t+1), \\
&\text{and } \mybf{W}_t = \mybf{Z}_t - \E\left(\mybf{Z}_t | X_{\lfloor 0,t \rfloor}\right)
\end{align*}

Independence of $X_{\lfloor 0,t \rfloor}$ and $\mybf{W}_{\lfloor 0,t \rfloor}$ holds thanks to the Gaussian assumption.

Let :\begin{align}
\label{e:lemm1}  f_t(X_{\lfloor 0,t \rfloor},\mybf{Z}_{\lfloor 0,t \rfloor}) & = f_t(X_{\lfloor 0,t \rfloor},\tilde{\mybf{X}}_{\lfloor 0,t \rfloor}+\mybf{W}_{\lfloor 0,t \rfloor})\\
& =  f_t(X_{\lfloor 0,t \rfloor},\Lambda X_{\lfloor 0,t \rfloor}+\mybf{W}_{\lfloor 0,t \rfloor})\\
& = g_t(X_{\lfloor 0,t \rfloor},\mybf{W}_{\lfloor 0,t \rfloor})
\end{align}

Let us now compute $\tilde{\mybf{X}}_{\lfloor 0,t \rfloor}$ defined as : 
\[\E(\mybf{Z}_{\lfloor 0,t \rfloor}|X_{\lfloor 0,t \rfloor})=\left\lbrace \E(\mybf{Z}_u|X_{\lfloor 0,t \rfloor}),\; 0 \leq u \leq t \right\rbrace = \tilde{\mybf{X}}_{\lfloor 0,t \rfloor}\]
 $(X_t,\mybf{Z}_t)$ being a Gaussian vector, conditional expectations with respect to $X_{\lfloor 0,t \rfloor}$ are the projections on the linear space generated by $X_{\lfloor 0,t \rfloor}$. 
 \begin{hyp}
 \label{h:full:rank}
For every $t$ we suppose that $\mybf{U}_{\lfloor 0,t \rfloor}$ is of full rank.
\end{hyp}
Thus : 
%$\left\lbrace X_s,\; 0\leq s \leq t \right\rbrace$
\begin{equation}
\E(Z_u^j|X_{\lfloor 0,t \rfloor})= \mybf{\lambda}^j_{\lfloor 0,t \rfloor,u} X^*_{\lfloor 0,t \rfloor}
\end{equation}
where $\mybf{\lambda}^j_{\lfloor 0,t \rfloor,u}$ is a vector of size $(t+1)$ given by classical linear regression results :
\begin{equation}
\label{e:lambda:calc}
\mybf{\lambda}^j_{\lfloor 0,t \rfloor,u}= \left(\Gamma^{XX}_{\lfloor 0,t \rfloor,\lfloor 0,t \rfloor}\right)^{-1} \gamma^{XZ^j}_{\lfloor 0,t \rfloor,u}
\end{equation}
$\Gamma^{XX}_{\lfloor 0,t \rfloor,\lfloor 0,t \rfloor}$ is invertible as consequence of assumption \ref{h:full:rank}. 

Let $\Gamma_{\lfloor 0,t \rfloor,\lfloor 0,t \rfloor}^{X,Z}$ defined as previously in (\ref{def:matrice:Gamma}) and 
\begin{equation}
\Lambda^{XZ}_{\lfloor 0,t \rfloor,\lfloor 0,t \rfloor}=\left(\Gamma^{XX}_{\lfloor 0,t \rfloor,\lfloor 0,t \rfloor}\right)^{-1} \Gamma^{XZ}_{\lfloor 0,t \rfloor,\lfloor 0,t \rfloor}
\end{equation} 
 then :
\begin{equation}
\label{e:lambda}
\tilde{\mybf{X}}_{\lfloor 0,t \rfloor}=\Lambda^{XZ}_{\lfloor 0,t \rfloor,\lfloor 0,t \rfloor} X_{\lfloor 0,t \rfloor}
\end{equation}
$\tilde{\mybf{X}}_{\lfloor 0,t \rfloor}$ is a $(p-1)\times (t+1)$ matrix as is $ \mybf{W}_{\lfloor 0,t \rfloor} = \mybf{Z}_{\lfloor 0,t \rfloor}-\tilde{\mybf{X}}_{\lfloor 0,t \rfloor}$.

Thanks to \eqref{e:lemm1} we have : 
\[ S^X_t = \frac{ \Var\left( \E\left(g_t(X_{\lfloor 0,t \rfloor},\mybf{W}_{\lfloor 0,t \rfloor})|X_{\lfloor 0,t \rfloor}\right)\right) }{\Var (Y_t)}. \]

Note that the space of all the square integrable functions of the form $\phi(X_{\lfloor 0,t \rfloor})$ is the same as the space of all the square integrable function of the form $\psi(X_{\lfloor 0,t \rfloor}, \tilde{\mybf{X}}_{\lfloor 0,t \rfloor})$.\\ Thus $\E(Y_t|(X_{\lfloor 0,t \rfloor}, \tilde{\mybf{X}}_{\lfloor 0,t \rfloor}))= \E(Y_t|X_{\lfloor 0,t \rfloor})$. For $t$ fixed, we are now exactly in the previous case of two groups of independent inputs $X_{\lfloor 0,t \rfloor}$ and $\mybf{W}_{\lfloor 0,t \rfloor}$ and thus we can apply the Pick and Freeze method \eqref{e:lemm1} and \eqref{e:pickfreeze} with $\mybf{X}=\left(X_{\lfloor 0,t \rfloor},\tilde{\mybf{X}}_{\lfloor 0,t \rfloor}\right)$ and $\mybf{Z}=\mybf{W}_{\lfloor 0,t \rfloor}$. By copying $\mybf{W}_t'$ of $\mybf{W}_t$ we mean a stochastic process independent of $\mybf{W}_t$ with the same finite dimensional distributions. If $(X_{\lfloor 0,t \rfloor}^{(i)},\mybf{W}_{\lfloor 0,t \rfloor}^{(i)})_{i=1,\dots,N}$ is a sample of $(X_{\lfloor 0,t \rfloor},\mybf{W}_{\lfloor 0,t \rfloor})_{t\in \N}$ we denote $(X_{\lfloor 0,t \rfloor}^{(i)} ,(\mybf{W}_{\lfloor 0,t \rfloor}^{(i)})')_{i=1,\dots,N}$ the sample obtained with $\mybf{W}'_{\lfloor 0,t \rfloor}$ a copy of $\mybf{W}_{\lfloor 0,t \rfloor}$.

Let $Y^X_t=g_t(X_{\lfloor 0,t \rfloor},\mybf{W}'_{\lfloor 0,t \rfloor})$. 

We have, for any $t\in\N$:
\[ S^X_t = \frac{\Cov(Y_t, Y_t^X)}{\Var (Y_t)}, \]
where $Y_t^X = f_t\left(X_{\lfloor 0,t \rfloor}, \tilde{\mybf{X}}_{\lfloor 0,t \rfloor}+\left(\mybf{W}_s'\right)_{s\leq t}\right)$,
where $X_{\lfloor 0,t \rfloor}$ and thus $\tilde{\mybf{X}}_{\lfloor 0,t \rfloor}$, which is a function of $X_{\lfloor 0,t \rfloor}$, are frozen.

Now to estimate $S^X_t$, we need only to get an empirical estimator of the covariance as in (\ref{e:pickfreeze}). Thus we simulate a sample $(X_{\lfloor 0,t \rfloor}^{(i)},\mybf{W}_{\lfloor 0,t \rfloor}^{(i)})$ and $(X_{\lfloor 0,t \rfloor}^{(i)}, (\mybf{W}_{\lfloor 0,t \rfloor}^{(i)})')$, $i=1,\dots, N$.

To do this, we simulate two independent pairs $(X_{\lfloor 0,t \rfloor}^{(i)},\mybf{Z}_{\lfloor 0,t \rfloor}^{(i)})$ and $((X_{\lfloor 0,t \rfloor}^{(i)})',(\mybf{Z}_{\lfloor 0,t \rfloor}^{(i)})')$. Thanks to these pairs, we build $\tilde{\mybf{X}}_{\lfloor 0,t \rfloor}^{(i)}$ and $(\tilde{\mybf{X}}_{\lfloor 0,t \rfloor}^{(i)})'$. We deduce $(\mybf{W}_{\lfloor 0,t \rfloor}^{(i)})'$ thanks to \\ $(\mybf{W}_{\lfloor 0,t \rfloor}^{(i)})'= (\mybf{Z}_{\lfloor 0,t \rfloor}^{(i)})' -(\tilde{\mybf{X}}_{\lfloor 0,t \rfloor}^{(i)})'$. 

In the stationary case, the simulation of the Gaussian vectorial process $(\mybf{U}_t)_{t\in\N}=(X_t,\mybf{Z}_t)_{t\in\N}$ is a classical problem when its covariance is known. In the non stationary case the Cholesky decomposition of the covariance matrix is the most popular method.
%The simulation of Gaussian processes is in stationary case is a classical problem with to main issues, frequented representation and use of FFT and use of  

Once $(\mybf{U}_t)_{t\in \N}$ is simulated, we have to recover $\tilde{\mybf{X}}_{u}$ and $\mybf{W}_{u}=\mybf{Z}_u-\tilde{\mybf{X}}_{u}$. Formula (\ref{e:lambda}) gives $\tilde{\mybf{X}}_{\lfloor 0, t \rfloor}= \Lambda^{XZ}_{\lfloor 0,t \rfloor,\lfloor 0,t \rfloor}X_{\lfloor 0,t \rfloor}$, and directly allows the computation of :\\$\mybf{W}_{\lfloor 0,t \rfloor}= \mybf{Z}_{\lfloor 0,t \rfloor}-\Lambda^{XZ}_{\lfloor 0,t \rfloor,\lfloor 0,t \rfloor}X_{\lfloor 0,t \rfloor}$.

\section{Sensitivity and stationarity}
\subsection{Stationary input-output models }
Let $\mybf{U}_t=(X_t,\mybf{Z}_t)_{t\in \Z}$ a stochastic process considered as an input and $Y_t=f_t(\mybf{U}_t,\dots,\mybf{U}_0)$ as the output.
It is assumed in the following that $(\mybf{U}_t)_{t\in\Z}$ is a stationary process.
Remember that a process is stationary if all its multidimensional distributions are translation invariant in time. For a Gaussian process $(\mybf{U}_t)_{t\in \Z}$, stationarity is equivalent to $\E(\mybf{U}_t)=m$ and $\E(\mybf{U}_t \mybf{U}^*_{t+k})= \Gamma^U(k)$ independents of $t$.

We consider two cases : \begin{itemize}
\item Case 1 : $( \mybf{U}_t)_{t\in\Z}$ is stationary and $Y_t= f(\mybf{U}_t,\dots,\mybf{U}_{t-M})$ is a stationary process $M$ is fixed as the proper memory of $Y_t$
\item Case 2 : $(\mybf{U}_t)_{t\in\Z}$ is stationary and there exists a stationary process $Y^{\star}_t$ (Bernoulli shift process) such as 
$Y^{\star}_t=f(\mybf{U}_t,\dots,\mybf{U}_{0},\mybf{U}_{-1},\dots)$ and
 $Y_t= f_t(\mybf{U}_t,\dots,\mybf{U}_{0})=f(\mybf{U}_t,\dots,\mybf{U}_{0},0,\dots)$.
\end{itemize}

In the second case, $Y^{\star}_t$ is a stationary process while $Y_t$ is not strictly stationary but it is a useful approximation in applications as we will see later.
\subsection{Sobol indices convergence}
We first study the case 1 $Y_t= f(\mybf{U}_t,\dots,\mybf{U}_{t-M})$. We assume $Y_t$ centered, without loss of generality.
The Sobol index is defined as :
\[S^{X}_t = \frac{\Var\left( \E\left(Y_t|X_{\lfloor 0,t \rfloor}\right)\right)}{\Var (Y_t)}\]

For a fixed $K$, let note $X_{\lfloor t-K, t \rfloor}= \lbrace X_t, \dots, X_{t-K} \rbrace$, we have :
\begin{equation}
\forall
 t \geq K \;\;\;
\frac{\Var \left( \E\left(Y_t|X_{\lfloor 0,t \rfloor}\right)\right)}{\Var (Y_t)} \geq 
\frac{\Var\left( \E\left(Y_t|X_{\lfloor t-K,t \rfloor}\right)\right)}{\Var (Y_t)} = S^X_{t,K} 
\end{equation}
But the last quantity is constant in $t$ by translation invariance, $(Y_t,\mybf{U}_t)$ being stationary thus : 
\begin{equation}
%\text{ For } t \geq K \;\;\;\; \forall u  \; \;\;S^X_{t+u,K}=S^X_{t,K} \Rightarrow S^X_{t,K} = S^X_{K} 
\text{ for } t \geq K  \;\;\; S^X_{t,K} = S^X_{K} 
\end{equation}
$S^X_{K}$ is an increasing sequence in $K$ and bounded so :
\begin{equation}
S^X_{\infty} = \sup_{K} S^X_K
\end{equation}

thus we see that the sensitivity reaches a limit $S_{\infty}^X$ as $t\rightarrow +\infty$ whatever the stationary system $(Y_t,\mybf{U}_t)$.
 \begin{lemma}{In Case 1 :}
 \begin{equation}
 \lim_{t\rightarrow +\infty} S^X_t= \lim_{t\rightarrow +\infty} \sup_{K\leq t } S^X_K = S_{\infty}^X 
 \end{equation}
 \end{lemma}

We give in appendix the proof of the same result in the second case when $Y_t$ has its proper dynamics, but only in particular cases when $\mybf{U}_t$ is a linear causal process and $Y_t$ has a specific form, including the most general linear case.

\subsection{$VAR$ input case}
The simplest input model is the vectorial autoregressive process of order $p$, noted $VAR(p)$.

The $VAR(1)$ model is given by :
\begin{eqnarray}
\label{autoReg}
\begin{pmatrix}
X_t\\
\mybf{Z}_t
\end{pmatrix}
&=&A\begin{pmatrix}
X_{t-1}\\
\mybf{Z}_{t-1}
\end{pmatrix}+
\mybf{\omega}_{t}, \;\;\; t>0
\end{eqnarray}
%and $X_{-1}=Z_{-1}=0$.
%$=(\omega_t^j)_{t\in\N, j=1,\ldots,p+q}$ 
$A$ is a $p \times p $ matrix, and $(\mybf{\omega}_t)_{t\in\N} $ are $p-$dimensional iid standard Gaussian variables with covariance $\Theta$ and $(X_0,\mybf{Z}_0)$ given.

We have of course:
\begin{equation}
\label{ARX}
\begin{pmatrix}
X_t\\
\mybf{Z}_t
\end{pmatrix}=
\sum_{k=0}^t A^k
\mybf{\omega}_{t-k}+ A^t\begin{pmatrix}
X_0\\
\mybf{Z}_0
\end{pmatrix} \\
\end{equation}

Stationarity is equivalent to a spectral radius $|\rho (A)| <1$. From now we suppose that this condition is verified. 
A $VAR(1)$ process is known to be geometrically ergodic (\cite{brockwell_time_2009}), thus whatever the distribution of $(X_0,\mybf{Z}_0)$ in (\ref{ARX}) is, it implies that when $t \rightarrow +\infty$ the distribution of $(X_t,\mybf{Z}_t)$ tends to the stationary distribution $\mathcal{N}(0,\Gamma_{t,t})$ with $\Gamma_{t,t}=\E((X_t,\mybf{Z}_t)(X_t,\mybf{Z}_t)^*)$ in the stationary case. We use this result in the following way : starting from any $(X_{t_1},\mybf{Z}_{t_1})$ for $t_1<0$  (for instance $t_1=-200$), the distribution of $(X_0,\mybf{Z}_0)$ is the stationary (invariant) distribution except negligible errors and using :
%Thus we can take for $\mybf{U}_0$ a random vector with the stationary (invariant) distribution and we can write :

\begin{equation}
\begin{pmatrix}
X_0\\
\mybf{Z}_0
\end{pmatrix}=
\sum_{k=t_1}^{0} A^{-k}
\mybf{\omega}_{k}+ A^{-t_1}\begin{pmatrix}
X_{t_1}\\
\mybf{Z}_{t_1}
\end{pmatrix} \\
\end{equation}

In stationary regime we have :
\begin{equation}
\label{ARX:stat}
\begin{pmatrix}
X_t\\
\mybf{Z}_t
\end{pmatrix}=
\sum_{k=0}^{\infty} A^k
\mybf{\omega}_{t-k}
\end{equation}

and thus : \begin{equation}
\Gamma_{t,t} = \sum_{k=0}^{\infty} A^k \Theta (A^k)^*
\end{equation}
$\Gamma_{\lfloor 0,t \rfloor,\lfloor 0,t \rfloor}^{X,X}$ can be easily inverted thanks to its Toeplitz property which allows a faster computation of $\Lambda_{\lfloor 0,t \rfloor,\lfloor 0,t \rfloor}^{XZ}$ (\ref{e:lambda:calc}) used for the simulation of $\tilde{\mybf{X}}$ (\ref{e:lambda}).

\begin{rque}
The $VAR(p)$ case, given by : 
\begin{eqnarray*}
\label{autoReg:p}
\begin{pmatrix}
X_t\\
\mybf{Z}_t
\end{pmatrix}
&=&\sum_{j=1}^p A^j\begin{pmatrix}
X_{t-j}\\
\mybf{Z}_{t-j}
\end{pmatrix}+
\mybf{\omega}_{t}, \;\;\; t>0
\end{eqnarray*}
can be reduced to the $VAR(1)$ case by increasing the dimension.
%of $ \begin{pmatrix}
%X_t\\
%\mybf{Z}_t
%\end{pmatrix}$.
\end{rque}

%%%%%%%%%%%%%%%%%
\begin{algorithm}                      % enter the algorithm environment
\caption{Reduction to independent inputs}          % give the algorithm a caption
\label{algo1}                           % and a label for \ref{} commands later in the document
\begin{algorithmic}[1]                    % enter the algorithmic environment
    \REQUIRE $A, \Theta, U, N, init_1, init_2$
    \STATE $times\leftarrow dim(U)[1]$ , $input\leftarrow dim(U)[2]$  
    \STATE $Simu_1[times,input,N] \leftarrow 0$, $Simu_2[times,input,N] \leftarrow 0$
    %, $Simu_{init}[times,input,n] \leftarrow 0$
    \FOR{$i=1$ to $input$}
   
        \STATE $\omega_1[times,input,N]\sim \mathcal{N}(0,\Theta)$ 
        \STATE $\omega_2[times,input,N]\sim \mathcal{N}(0,\Theta)$
        %\STATE $\omega_3[times,input,n]\sim \mathcal{N}(0,\Theta)$
       
        \STATE
        %\FOR{$t=2$ to $times$}
       % 	\STATE $Simu_{init}[t,,]\leftarrow A\cdot{}Simu_{init}[t-1,,] + %\omega_3[t,,]$ 
%        \ENDFOR
 %       \STATE
%        \STATE  \COMMENT{ Simulations of inputs process $U$}
%      	\STATE   $Simu_1[1,,]\leftarrow Simu_{init}[46,,]$	
%      	\STATE   $Simu_2[1,,]\leftarrow Simu_{init}[84,,] $  %\COMMENT{initialisation of $Simu_2$}
      	\STATE $Simu_1[1,,] \leftarrow  init_1 $
      	\STATE $Simu_2[1,,] \leftarrow  init_2 $
      	
      	\FOR{$t=2$ to $times$}  
        	\STATE $Simu_1[t,,]\leftarrow A \cdot{}Simu_1[t-1,,] + \omega_1[t,,]$ 
        	\STATE $Simu_2[t,,]\leftarrow A\cdot{}Simu_2[t-1,,] + \omega_2[t,,]$ 
        \ENDFOR       
        \STATE
        
        \FOR{$t=1$ to $times$}
	        \STATE $\Lambda\leftarrow (\Cov(U[1:t,i],U[1:t,i]))^{-1} \Cov(U[1:t,],U[1:t,])$
    	    \STATE $ \tilde{X}_1[t,,] \leftarrow (Simu_1[1:t,i,])^* \cdot{}\Lambda $ 
        	\STATE $\tilde{X}_2[t,,]\leftarrow (Simu_2[1:t,i,])^* \cdot{}\Lambda$
		\ENDFOR

        \STATE $W_1\leftarrow Simu_1-\tilde{X}_1$
        \STATE $W_2\leftarrow Simu_2-\tilde{X}_2$
        
        \STATE $SIMU_1 \leftarrow Simu_1$
        \STATE $SIMU_2 \leftarrow \tilde{X}_1+ W_2 $
    \ENDFOR
    \RETURN $(SIMU_1, SIMU_2)$
\end{algorithmic}
\end{algorithm}
%%%%%%%%%%%%%%%%%%%%
A pseudo code regarding the reduction of dependent inputs to independent inputs is presented in algorithm \ref{algo1} for $VAR(1)$. The algorithm of the Pick and Freeze method is presented in algorithm \ref{algo2}. 
%The pseudo code is written for causal dependent inputs $\left(\mybf{U}_t\right)_{t\in \N}=\left(\left(X_t,\mybf{Z}_t\right)\right)_{t\in\N}$, defined as in equation (\ref{autoReg}).
 We build samples of size $N$.\\
%The initialisation lines $(12)$ and $(13)$, in the algorithm \ref{algo1}, are here to ensure that processes
We have to pay attention as explained above to the initial value that we choose for processes $Simu_1$ and $Simu_2$ to ensure that they are stationary. To do that we can use a process $Simu_{init}$, with some initialisation, that we simulate.  
We select at a time $t_1$, $Simu_{init}[t_1]$ that will be the initialisation.\\
Lines (17) and (18), $W_1$ and $W_2$ are what we call previously $W$ and $W'$. The same for $Y_1$ and $Y_2$ are what we call previously $Y$ and $Y'$ \\
Algorithm \ref{algo1} returns $SIMU_1$ and $SIMU_2$ which is its pick-freezed replication.  They will be used to estimate the Sobol index in algorithm \ref{algo2}.

\begin{algorithm}                      % enter the algorithm environment
\caption{Pick and Freeze estimation}          % give the algorithm a caption
\label{algo2}                           % and a label for \ref{} commands later in the document
\begin{algorithmic}[1]                    % enter the algorithmic environment
    \REQUIRE $SIMU_1, SIMU_2, init$
    \STATE $times\leftarrow dim(SMU_1)[1]$ , $input\leftarrow dim(SIMU_1)[2]$  
    \STATE $indice \leftarrow 0$
        
    \FOR{$i=1$ to $input$}
    	\IF{ init $\neq$ NULL}
	    	\STATE $Y_1 = f(SIMU_1, init)$
    		\STATE $Y_2 = f(SIMU_2, init)$
    	\ELSE
    		\STATE $Y_1 = f(SIMU_1)$
    		\STATE $Y_2 = f(SIMU_2)$
    	\ENDIF
    	 
	    \STATE $ indice[,input] = \frac{\E( Y_1 \otimes Y_2) - \E(Y_1) \E(Y_2) }{\Var(Y_1)} $
    \ENDFOR    
    \RETURN $indice$
\end{algorithmic}
\end{algorithm}

\begin{rque}
Algorithm \ref{algo1} and  \ref{algo2} can be applied to any gaussian stationary process. Part 1 of algorithm \ref{algo1} is a method of simulation of the process. 
\end{rque}
\subsection{Toy models}

We study two stationary toy models one linear and one non linear given by: 
\begin{align}
\label{model:jouet:1}
& Y_t=0.2 Y_{t-1} + 0.3 X_t + Z_t  \\
\label{model:jouet:2}
& Y_t = X_t Z_t + 0.2exp(-Z_t) 
\end{align}

$\begin{pmatrix}
X_t\\
Z_t
\end{pmatrix}$ is a $VAR(1)$ stationary process given by:
\begin{eqnarray}
\begin{pmatrix}
X_t\\
Z_t
\end{pmatrix}
&=&
\begin{pmatrix}
0.8 & 0.4\\
0.1   & 0.2
\end{pmatrix}
\begin{pmatrix}
X_{t-1}\\
Z_{t-1}
\end{pmatrix}+
\omega_{t}
\end{eqnarray}

where $\omega_t$ a stationary Gaussian noise of covariance matrix $ \Theta = \begin{pmatrix}
0.1 & 0\\
0   & 0.1
\end{pmatrix}$.\\
%All the programs are done under R.

First let's give an example of the result that can be expected with the algorithm \ref{algo1} when we study the sensitivity of $Y$ with respect to $X$ for the model (\ref{model:jouet:1}). $\Lambda$ is here a vector $(\lambda_{t},\lambda_{t-1},\dots,\lambda_{t-k},\dots,\lambda_{0})^*$. Its values are given in table : \ref{t:lambda1}. After simulation of $Simu_1$ (table : \ref{t:simu1}), $\tilde{X}_1$ is calculated. Using the value of $\Lambda$ given in table \ref{t:lambda1} we compute the values corresponding to $\tilde{X}$ define as previously. The table \ref{t:simu1} is the result of the step (17) : $W_1=Z - \tilde{X}_1$. We do the same work with $Simu_2$ and we obtain $W_2$. We can remark that coefficients of $\Lambda$ decrease. Only the three first past instant are important. We see from table \ref{t:lambda1} that.

\begin{table}[H]
\begin{center}
\begin{tabular}{c||c c||c c||c c||c c||c c||}
 
\multicolumn{1}{c|}{\backslashbox{time}{time}} & \multicolumn{2}{c|}{0}   & \multicolumn{2}{c|}{1}& \multicolumn{2}{c|}{2}&\multicolumn{2}{c|}{3}&\multicolumn{2}{c|}{4}\\
\hline
0&    $X_t$  &0.12& $X_{t-1}$ & 0.38 & $X_{t-2}$ & 0.07 & $X_{t-3}$ &-0.01& $X_{t-4}$ &-0.01 \\
1&    &    & $X_{t}$ &-0.21 &$X_{t-1}$ &  0.33& $X_{t-2}$ &0.07 & $X_{t-3}$ &0.00  \\
2&    &    &   & 	   &$X_{t-1}$ &-0.21 & $X_{t-1}$ & 0.33& $X_{t-2}$ & 0.07 \\
3&    &    &   &      &  &		 & $X_{t}$  &-0.21 & $X_{t-1}$ &0.33  \\
4&    &    &   &      &  &		 &    &      & $X_{t}$ & -0.21 \\ 
\end{tabular}
\caption{Estimated values of $\Lambda$ (Step (13) of algorithm \ref{algo1})}
\label{t:lambda1}
\end{center}
\end{table}

\begin{table}[H]
\begin{center}
\begin{tabular}{c|c|c|| c | c |}
time & $X_t$ & $Z_t$ & $\tilde{X}_t$ & $W_t$ \\ \hline
0 & -0.21 &  0.40 & -0.02& 0.42\\
1 & 0.11  &-0.12 & -0.10 & -0.02\\
2 & -0.17 & - 1.02 & 0.06 & -1.08\\
3 & -0.29 & -0.79 & 0.01 & -0.80\\
4 &  0.24 & -0.80 & -0.16 & -0.64\\
\end{tabular}
\caption{Values of $X$, $Z$,  $\tilde{X}_t$ and $W$ for $Simu_1$ (Step (8) of algorithm \ref{algo1})}
\label{t:simu1}
\end{center}
\end{table}

\begin{figure}
   \begin{minipage}[c]{.46\linewidth}
 \includegraphics[scale=0.5]{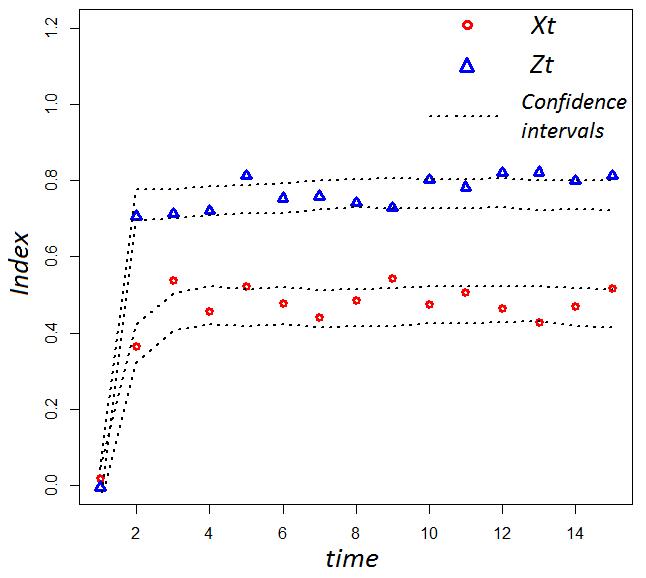}
 \caption{Toy model \ref{model:jouet:1} : Sobol index estimate in function of time. Sample size: 200}
\label{figure:model:jouet:1:200}  
   \end{minipage} \hfill
   \begin{minipage}[c]{.46\linewidth}
\includegraphics[scale=0.5]{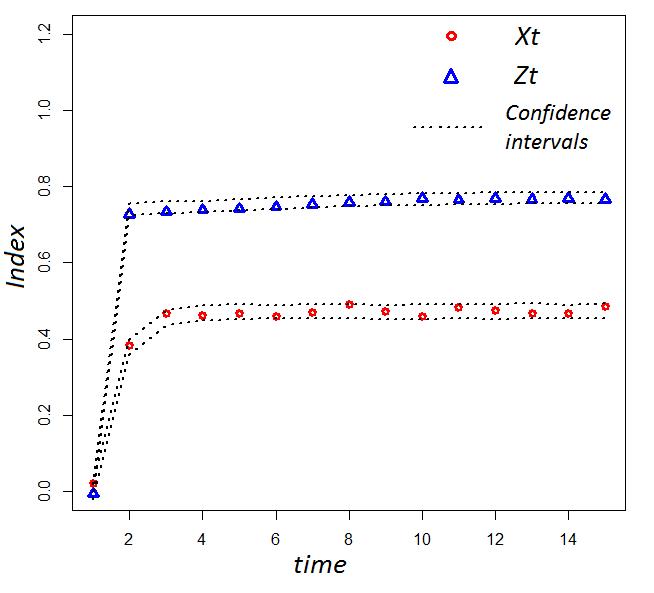}
 \caption{Toy model \ref{model:jouet:1} : Sobol index estimate in function of time. Sample size: 10000}
\label{figure:model:jouet:1:10000}
   \end{minipage} \hfill
\begin{minipage}[c]{.46\linewidth}
 \includegraphics[scale=0.5]{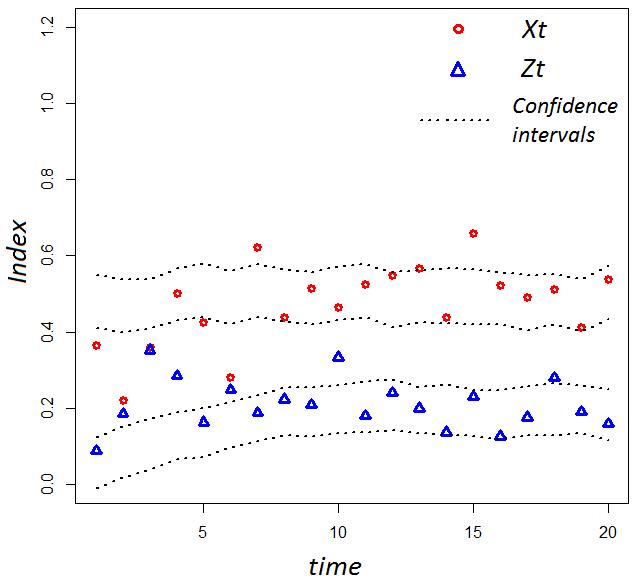}
 \caption{Toy model \ref{model:jouet:2} : Sobol index estimate in function of time. Sample size: 200}
\label{figure:model:jouet:2:200}
\end{minipage}\hfill
\begin{minipage}[c]{.46\linewidth}
 \includegraphics[scale=0.5]{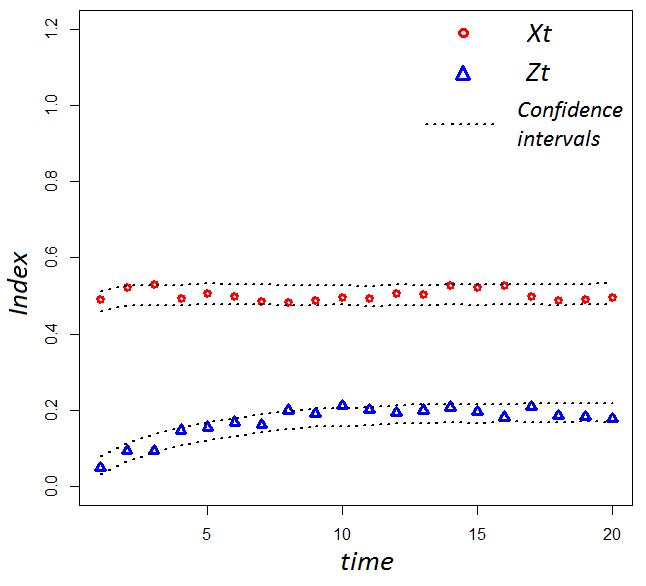}
 \caption{Toy model \ref{model:jouet:2} : Sobol index estimate in function of time. Sample size: 10000}
\label{figure:model:jouet:2:1000}
\end{minipage}    
\end{figure}   

Figures \ref{figure:model:jouet:1:200}, \ref{figure:model:jouet:1:10000}, \ref{figure:model:jouet:2:200}, \ref{figure:model:jouet:2:1000}, display for each time step, $S_t^X = \frac{\Var(\E(Y_t|X_{\lfloor 0,t \rfloor}))}{\Var(Y_t)}$, calculated for models (\ref{model:jouet:1}) and (\ref{model:jouet:2}) for different sizes of samples ($N=200$ and $N=10 000$). Confidence intervals are stated at the 95\% confidence level and plotted on each figures.
The following examples exhibit two different types of convergence :
\begin{itemize}
\item the convergence of the estimator. At each time step, the Pick and Freeze algorithm estimates $S_t^X$. The quality of the estimator $\widehat S^X_t$ depends on the size of the sample. The confidence interval is smaller when $N=10 000$. The convergence speed of the estimator is slow ($O(1/\sqrt{N})$) when we use a Monte Carlo sample. We could improve the speed using Quasi Monte Carlo (QMC) sample method \cite{saltelli2000sensitivity}. But in our case QMC seems hard to implement.

\item the temporal convergence : $\widehat S^X_t$ changes over time at each $t$. After just three iterations in time of the estimation algorithm, $S^X_t$ value reaches a limit (see on figures :  \ref{figure:model:jouet:1:200} and  \ref{figure:model:jouet:1:10000}).
Model \eqref{model:jouet:1} is auto-regressive, it means that $Y_t$ depends on its past $Y_{t-1}$ and so, on all the past of $X_t, Z_t$. We can rewrite it as :
\[ Y_t= \sum_{k=0}^{\infty} (0.2)^k (0.3 X_{t-k} + Z_{t-k}) \]
At time $t=0$ for example, $S^X_0= \frac{\Var(\E(Y_0|X_0))}{\Var(Y_0)}$. $Y_0$ is projected on a space of dimension 1 whereas it depends on $(X_{-\infty}, \dots, X_0, Z_{-\infty}, \dots, Z_{0})$. The projection space is too small. When increasing this space, the index increases and converges to a constant. It's what we call the phenomenon of memory which may refer to the physical concept of inertia. When $Y_t= \phi(X_t,Z_t)$ the index converges instantaneous (figures : \ref{figure:model:jouet:2:200}, \ref{figure:model:jouet:2:1000}).

The time convergence is interesting from a computing point of view. As the index converges to a constant, it useless to compute the index that takes into account the total trajectory of the process considered. It represents a lot of economy on a computational point of view because it requires less iterations for the evaluation of the index.
\end{itemize}
We can also notice that the method is independent of the expression of the model $f$. It is a black-box method requiring just the possibility to simulate a lot of inputs and outputs.

\section{Application}
\subsection{The physical problem}
%%C As tu des refs pour ce modele ? Meme un modele ressemblant...
%%C Est ce que la note de bas de page est vraie ?
%%C Est ce que Tint(0)=Tint(1)=0 est correct ?
 We now address a model of building constructed as a metamodel \cite{faivre2013analyse} estimated thanks to observed data. This building is actually a classroom that welcomes students during the school year. It is therefore sensitive to hours, days but also holidays that occur during the year.
 % The variable of interest and therefore the output of our model is the temperature inside the room. It is denoted by $T_ {int}$.
 
The studied room, whose internal temperature is $\Tint$ is surrounded by a corridor, an adjacent office, an office that is located below and a skylight (a shed). We have equipped these parts with temperature sensors. Temperatures are measured every hour denoted $(\Tbelow,\Tabove, \Toff, \Tcor ,\Text, \Tint) $. We also measure the temperature outside $\Text$. Obviously $\Text$ influences all the other temperatures $\Tbelow,\Tabove, \Toff, \Tcor, \Tint$.\\
 
We choose to study the summer period when students are not present and when the heating is off.

\subsection{The input and input-output models}
We have to build a statistical input model and also an input-output model. After preprocessing inputs are modelled as a $VAR(p)$ process. Details are given in appendix \ref{app:B}.

The output of the model is the internal temperature $\Tint$. Our sensitivity analysis aims to determine which among  $\Tbelow,\Tabove, \Toff, \Tcor ,\Text$ 
impact the most the variance of the internal temperature $\Tint$. 
%Due to cyclo-stationarity, we build a model for the day (8am-18pm) and night (19pm-7am) and switch model linking the day and night models. 
We choose to model the input-output system by a linear auto-regression of $\Tint$ on variables $U_t=(T^{e'}_t, T^{e'}_{t-1})_t$, for $e'\in \mathcal E' = \{ , \mathrm{below}, \mathrm{above}, \mathrm{off}, \mathrm{cor},\mathrm{ext}\}$. The used model is given in appendix \ref{app:B}.
Note that if we have at our disposal a physical model able to simulate the input-output model (for instance an electrical model), we can use the data provided by this model to apply the Pick and Freeze method.

\subsection{Numerical results}

\begin{figure}
\begin{center}
\includegraphics[scale=0.55]{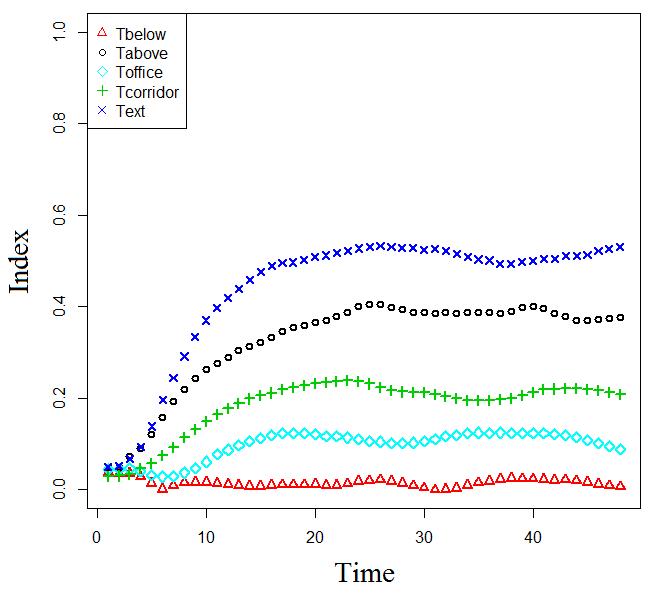}
\caption{Plots of estimated Sobol indices}
\label{f:superplot}
\end{center}
\end{figure}

We then compute the different estimators $\widehat S^X_t=\widehat S^e_t$, for $t=2,\ldots,48$ and $e\in \mathcal E'$. The results are gathered in figure : \ref{f:superplot}. Indices are estimated thanks to a sample of size $N= 5000$. 

The most influent variables are $\Text$ the external temperature and $\Tabove$ the temperature of the room above. These two rooms have the most important noise variances : 1.34 and 0.62 (see $\Theta$ matrix \eqref{e:theta}), whereas other noise variances are around 0.02. It seems logical that they are the most important variables in the sensitivity analysis. The number of time iterations is 20. The memory or inertia of the system is long enough.

Despite its small noise variance, $\Tcor$ is important. This variable illustrates the fact that sensitivity in a dependent context is not due only to the variance of the variable. Sensitivity is a function of the input-output relation and of the covariance between the variable of interest and the others.
$\Tcor$ is the coldest room, it lowers the temperature $\Tint$. $\Tcor$ is most likely to cause temperature variations. 

$\Tbelow$ is measured in a room located below the studied room $\Tint$ isolated by a slab. It seems logical that it does not affect $\Tint$.

\section*{Conclusion}
In this article, we have proposed a modified definition of Sobol indices, adapted to dependent dynamic inputs. The index is no longer associated with the Hoeffding decomposition and the sum of the index is not equal to 1. Nevertheless, the index is between 0 and 1 and keeps the same interpretation.
In a dynamic context the index varies with time but it is calculated as in a static case, that is to say that the input and the output are frozen at the same $t$ moment. With our proposition, we set in relief the dynamic aspect of the system by taking into account all the past of the input variables.  

To estimate this index in the dependent framework we have chosen a Pick and Freeze method because it is flexible : it works whatever the nature of the input (dynamic or static) and whatever the number of inputs. The only problem is that it requires independent inputs. 
Yet, we can use  this method even when the inputs are dependent and Gaussian because we can separate the inputs into two variables : one corresponding to the variable of interest (the one we need to study the sensitivity of the model) and another one which is totally independent. We can then apply the Pick and Freeze method on these new variables. We propose an algorithm to separate the variables and to calculate the index.

In the case of stationary variables this index approaches a limit quickly. On a computational perspective, this allows us to reduce the computation time, by only simulating the first instants of the process.
This method requiring to calculate the inverse covariance matrix of the inputs, $VAR$ inputs have several advantages. Their covariance matrix can be calculated analytically and can be inversed easily thanks to their Toeplitz properties. These processes are quickly simulated; this  is an advantage when we use the Pick and Freeze method which requires many input samples.

\appendix

\begin{appendices}
\section{Convergence of $S^X_t$ in the case where $Y_t=f_t(\mybf{U}_t,\dots,\mybf{U}_0)$\label{app:covergence}}

\begin{hyp}
$(\mybf{U}_t)_{t\in \Z}$ is a causal regular linear process. 
\end{hyp}

Causality means that $\mybf{U}_t$ does not depend on a distant past. Formally if $H^U_s$ is the Hilbert space (for the covariance scalar product) generated by $\lbrace \mybf{U}_v, v\leq s \rbrace$ then $\underset{s\in \Z}{\cap} H_s^U =\lbrace 0 \rbrace$. Causality and linearity are equivalent to the existence of a representation : 
\begin{equation}
\label{e:ma}
\mybf{U}_t= \sum_{k=0}^\infty C_k \mybf{\omega}_{t-k} 
\end{equation}
where $\lbrace \mybf{\omega}_t \rbrace$ is a vectorial white noise, $\E(\mybf{\omega}_t \mybf{\omega}^*_t) = \Theta $ and $C_k$ are $(p\times p)$ matrices such that $\Sigma \|C_k\| ^2 < \infty $.

Basic examples are $VARMA$ processes associated to recurrence equations : 
\[ P(d)(\mybf{U}_t)= Q(d) (\mybf{\omega}_t) \]
where $d$ is the backward operator defined as $d(\mybf{U}_t)= \mybf{U}_{t-1}, d(\mybf{\omega}_t)= \mybf{\omega}_{t-1}$ and $P,Q$ are polynomials such that $|P(z)|\neq 0$ for $|z| < 1$ for stationarity and $|Q(z)|\neq 0$ for regularity ($H^{\omega}_s=H^{U}_s$ for every $-\infty< s < +\infty$).

We need the following truncation result. Let $\eta$ fixed. We can find $K$ such that :
\[ \|\mybf{U}_t - \mybf{U}^{\tau(K)}_t \|\ < \eta \]
where $\mybf{U}^{\tau(K)}_t= \sum_{k=t}^{t-K} C_k \mybf{\omega}_{t-k}$
 \[ \E \left( | \sum_{k=t-K+1}^{\infty} C_k \mybf{\omega}_{t-k}|\right) \leq \sum_{k=K+1}^{\infty} \| \Theta \| \| C_k\|^2 \]

\begin{hyp}{on the output}
\begin{itemize}
\item $Y^{\star}_t$ has the denominated truncation property as $t\rightarrow \infty$ : \[\lim\limits_{t \to \infty} \E( |Y^{\star}_t-Y_t|^2)=0\]  and \[ \lim\limits_{t \to \infty} |E(|Y^{\star}_t|^2)- E(|Y_t|^2)| = 0\]
\item $Y^{\star}_t$ has the conditional past truncation property as $t\rightarrow \infty$ : 
\[\lim\limits_{t \to \infty} \E( \E(Y^{\star}_t | X_{\lfloor -\infty, t \rfloor})^2 ) -\E( \E(Y^{\star}_t | X_{\lfloor 0, t \rfloor})^2 ) = 0\] 
\end{itemize}
\end{hyp}

From the Jensen inequality :
\[ \E\left( \E(Y^{\star}_t | X_{\lfloor 0, t \rfloor})- \E(Y_t | X_{\lfloor 0, t \rfloor}) \right)^2 \leq  \E(Y^{\star}_t-Y_t)^2 \] tends to zero as $t \rightarrow \infty$

$\E\left( \E(Y^{\star}_t | X_{\lfloor -\infty, t \rfloor})^2 \right)$ is constant by translation invariance and denoted $(S^X_t)^{\star}. \Var(Y_t)$.

From assumption 2 $\E( \E(Y^{\star}_t | X_{\lfloor 0, t \rfloor})^2 ) \rightarrow (S^X_t)^{\star}. \Var(Y_t)$ as $t \rightarrow \infty$

By assumption 1 :

We can rewrite : 
\[ \E(Y_t|X_{\lfloor 0, t \rfloor})= \E(Y^{\star}_t|X_{\lfloor 0, t \rfloor})+\E(Y_t-Y^{\star}_t|X_{\lfloor 0, t \rfloor}) \]

Then :
\[ \E(\E(Y_t|X_{\lfloor 0, t \rfloor})^2)=\E( \E(Y^{\star}_t|X_{\lfloor 0, t \rfloor})^2)+\E(\E(Y_t-Y^{\star}_t|X_{\lfloor 0, t \rfloor})^2)+2\E( \E(Y^{\star}_t|X_{\lfloor 0, t \rfloor})\E(Y_t-Y^{\star}_t|X_{\lfloor 0, t \rfloor})) \]

When $t$ tends to $+\infty$ :
\begin{align*}
& \lim\limits_{t \to \infty} \E( \E(Y^{\star}_t|X_{\lfloor 0, t \rfloor})^2)=(S^X_t)^{\star}. \Var(Y_t) \\
& \lim\limits_{t \to \infty} \E(\E(Y_t-Y^{\star}_t|X_{\lfloor 0, t \rfloor})^2) = 0 \\
& \lim\limits_{t \to \infty} \E( \E(Y^{\star}_t|X_{\lfloor 0, t \rfloor})\E(Y_t-Y^{\star}_t|X_{\lfloor 0, t \rfloor})) =0 \text{ from Schwartz ineguality}\\
\end{align*}

So under the hypotheses 1 and 2 : \begin{equation}
\lim\limits_{t \to \infty} S^X_t = (S^X_t)^{\star}
\end{equation}

\paragraph{Example :}
Let us check assumptions 1 and 2 for the example :
\[Y_t^{\star}= \alpha Y_{t-1}^{\star}+ B \mybf{U}_t \]
for $\mybf{U}_t$ causal regular Gaussian process and $|\alpha|<1$.\\
$Y_t$ can be written as :
\[Y_t^{\star}= \sum_{k=0}^{\infty} \alpha^k B \mybf{U}_{t-k} \] and \[Y_t= \sum_{k=0}^{t} \alpha^k B \mybf{U}_{t-k} \] 
thus :
\[ \E|Y_t^{\star}-Y_t|^2 \leq C |\alpha|^{2t} \] thus the assumption 1 is verify.
Now we calculate :
\[ \E( Y_t^{\star} | X_{\lfloor -\infty,t \rfloor})= \sum_{k=0}^{\infty} \alpha^k B \E(\mybf{U}_{t-k} | X_{\lfloor -\infty,t \rfloor}) \]

we know that \[ \E(\mybf{U}_{t-k} | X_{\lfloor -\infty,t \rfloor})=\sum_{j=0}^{\infty} C_j \E(\omega_{t-j}| X_{\lfloor -\infty,t \rfloor})\]
and if $\eta$ given it exists $K$. \\
$ \| U_{t-k} - U^K_{t-k} \| \leq \eta$  implies $ \| \E( U_{t-k}) - \E(U^K_{t-k}) \| \leq \eta$
 for $K$ large enough from the truncation property of causal processes. But from the regularity of $\mybf{U}_t$ we have :$H^X_u=H^{\omega}_u$ for all $u$ thus in Gaussian case :
$\E( \omega_{t-j}|X_{\lfloor -\infty,t \rfloor})= \E( \omega_{t-j}|\omega_{\lfloor -\infty,t \rfloor})$.
But for $t-j>0$, $\omega_{t-j}$ is independent of $\lbrace \omega_s, s<0 \rbrace$.

Thus \begin{align*}
\E(\omega_{t-j}|X_{\lfloor -\infty,t \rfloor}) & = \E(\omega_{t-j}|\omega_{\lfloor -\infty,t \rfloor})\\
& = \E(\omega_{t-j}|\omega_{\lfloor 0,t \rfloor})\\
& = \E(\omega_{t-j}|X{\lfloor 0,t \rfloor})
\end{align*}
and we have proved the result choosing $t$ large enough to have $t-K>0$.
\section{Construction of the statistical metamodel \label{app:B}}
To build a model easy to use for simulations, we need to preprocess the data in order to be placed in a stationary condition. We built a model for summers. $T^e_t$ is a scalar time series of input data. Preprocessing means first to transform $\bar{T}^e_t$ into $T^e_t$ :

\[ \bar{T}^e_{t}=S(t)+V(t)*T^e_{t} \;\; \forall e\in \mathcal E \] 
where $\mathcal E = \{ \mathrm{int}, \mathrm{ext}, \mathrm{above}, \mathrm{below}, \mathrm{cor}, \mathrm{off}\}$,
 $S(t)$ is the mean with period 24 hours, $V(t)$ the variance periodic function and $T^e_{t}$ a stationary or cyclo-stationary process. 
 
%%%%%%%%%%%%%%%%%%%%%%%%%%%%%%%%%%%%%%%%%%%%%%%%%%%%%%
Temperatures  ($\Tbelow,\Tabove, \Toff, \Tcor ,\Text, \Tint$) are modelled by a $VAR$ process for working days:
%with an exogenous variable $N$ of order 1. 
\[ \forall e \in \mathcal E , \;\;
T_t^{e} = \sum_{f \in \mathcal E } \gamma_f^e T_{t-1}^f +\omega_t^{e}, \]
where $(\omega_t^{e})_{t\in\N, e\in \mathcal E}$ are Gaussian variables with covariance $\Theta$.
%For night model, $N=0$.
%We model the cyclo stationary process $N$ by three $AR(1)$ process with a Gaussian white noise. For days/working day : $ N_t = \alpha N_{t-1} +\varepsilon_t \;\; \alpha>0 $ and for night : $N_t=0 $.
Let $\mybf{T}_t=(T^e_t)_{e\in \mathcal{E}} , \;\; \mybf{\omega}_t=(\omega_t^{e})_{e\in \mathcal{E}}$ thus :
\[ \mybf{T}_t = \sum_{l=1}^p D_l \mybf{T}_{t-p} + \mybf{\omega}_t \]
For every fixed $p$ we estimated $(D_l, \;\; l\leq 1,\dots, p )$ and $\Theta$ the covariance of $\mybf{\omega}$ by maximum of likelihood under the constraint imposed by the stationarity of $\mybf{T}_t$. Then we choose $p$ using AIC criteria; obtaining $p=2$.

$D_1$ and $D_2$ estimated are :
 {\footnotesize \begin{equation*}
\begin{array}{c c}
D_1=
\begin{pmatrix}
0.88 & 0.01 & 0.06 & 0.06 & 0 \\
-0.23 & 1.21 & 0.01 & 0.49 & 0.3 \\
0.21  & 0.03 & 1.24 & 0.06 & 0.01 \\
0.12 & 0.06 & 0.03 & 1 & 0.01 \\
0.20 & 0.61 & -0.19 & 0.71 & 0.85
\end{pmatrix} & 
D_2= 
\begin{pmatrix}
0.04 & -0.01 & -0.07 &-0.04 & 0.01 \\
0.03 & -0.48 & -0.04 & -0.08 & -0.04 \\
-0.16 & -0.02 &-0.30 & -0.09 & -0.10 \\
-0.03 & -0.04 & -0.05 & -0.15 & -0.01\\
-0.45 & -0.60 & 0.25 & -0.31 & -0.05 
\end{pmatrix}
\end{array}
\end{equation*}}
The $\Theta$ matrix estimated is : 
 {\footnotesize
\begin{equation}
\label{e:theta}
\Theta = \begin{array}{l}
\Tbelow \\
\Tabove \\
\Toff \\
\Tcor \\
Text
\end{array} \begin{pmatrix}
0.01 & 0.01 &0.004 & 0.01 & 0.01 \\
0.01 & 0.62 & 0.02 & 0.03 & 0.46 \\
0.004 & 0.02 & 0.03 & 0.002 & 0.04 \\
0.01 & 0.03 & 0.002 & 0.02 & 0.04 \\
0.01 & 0.46 & 0.04 & 0.04 & 1.34
\end{pmatrix}
\end{equation}}

The input-output system is modelled by an auto-regression of $\Tint$ on variables \\ $\mybf{U}_t=(\Tbelow,\Tabove, \Toff, \Tcor ,\Text)$ :
\begin{equation}
\label{e:supermodel}
\forall t \geq 1, \;\; \Tint(t) = \sum_{e\in\mathcal E, k=1}^2 \phi_{e,k} T_{e,t-k} + \sum_{k=1}^2 \phi_{int,k} \Tint_{t-k} 
\end{equation}

%\begin{align}
%\forall t \geq 1, \;\; \Tint(t)& = \Tint( \Tabove, \Tbelow, \Tcor, \Toff,\Tint,\Text)\\
%\label{e:supermodel}
% & = \sum_{e\in\mathcal E, k=1}^2 \phi_{e,k} T_{e,t-k} + \sum_{k=1}^2 \phi_{int,k} \Tint_{t-k} 
%\end{align}

where $\mathcal E = \{ \mathrm{int}, \mathrm{above}, \mathrm{below}, \mathrm{cor}, \mathrm{off}, \mathrm{ext}, \}$

\end{appendices}
\section*{Acknowledgements}
The authors would like to thank the reviewers for their helpful comments.
%\newpage
\iffalse
\nocite{mara2011variance}
\nocite{xu2008uncertainty}
\nocite{tissot2012bias}
\nocite{owen2005multidimensional}
\nocite{luenberger1968optimization}
\nocite{kucherenko2012estimation}
\nocite{jacques2006sensitivity}
\nocite{ratto2007state}
\nocite{box1987empirical}
\nocite{marrel2008efficient}
\nocite{williams2006gaussian}
\nocite{marrel2009calculations}
\nocite{racine1993efficient}
\nocite{lauret2006node}
\nocite{tarantola2006random}
\nocite{chastaing2012generalized}
\nocite{da2009local}
\nocite{eubank_nonparametric_1999}
\nocite{brockwell_time_2009}
\nocite{saltelli_sensitivity_2004}
\nocite{saltelli_sensitivity_2008}
\nocite{fort2013estimation}
\fi 
\bibliographystyle{unsrt}
\bibliography{references,biblio}

\end{document}